\documentstyle[aps,prl,epsf]{revtex}
\begin{document}                                             
\begin{center}

{\bf \Large
Search for a Technicolor $\omega_T$  Particle \\
in Events with a Photon and a $b$-quark Jet at CDF 
}

\font\eightit=cmti8
\def\r#1{\ignorespaces $^{#1}$}
\hfilneg
\begin{sloppypar}
\noindent
F.~Abe,\r {17} H.~Akimoto,\r {39}
A.~Akopian,\r {31} M.~G.~Albrow,\r 7 A.~Amadon,\r 5 S.~R.~Amendolia,\r {27} 
D.~Amidei,\r {20} J.~Antos,\r {33} S.~Aota,\r {37}
G.~Apollinari,\r {31}
% ...
\end{sloppypar}
\vskip .026in
\begin{center}
(CDF Collaboration)
\end{center}

\vskip .026in
\begin{center}
\r 1  {\eightit Argonne National Laboratory, Argonne, Illinois 60439} \\
\r 2  {\eightit Istituto Nazionale di Fisica Nucleare, University of Bologna,
I-40127 Bologna, Italy} \\
\r 3  {\eightit Brandeis University, Waltham, Massachusetts 02254} \\
\r 4  {\eightit University of California at Los Angeles, Los 
Angeles, California  90024} \\  
\r 5  {\eightit University of Chicago, Chicago, Illinois 60637} \\
\r 6  {\eightit Duke University, Durham, North Carolina  27708} \\
\r 7  {\eightit Fermi National Accelerator Laboratory, Batavia, Illinois 
60510} \\
\r 8  {\eightit University of Florida, Gainesville, Florida  32611} \\
\r 9  {\eightit Laboratori Nazionali di Frascati, Istituto Nazionale di Fisica
               Nucleare, I-00044 Frascati, Italy} \\
\r {10} {\eightit University of Geneva, CH-1211 Geneva 4, Switzerland} \\
\r {11} {\eightit Harvard University, Cambridge, Massachusetts 02138} \\
\r {12} {\eightit Hiroshima University, Higashi-Hiroshima 724, Japan} \\
\r {13} {\eightit University of Illinois, Urbana, Illinois 61801} \\
\r {14} {\eightit Institute of Particle Physics, McGill University, Montreal 
H3A 2T8, and University of Toronto,\\ Toronto M5S 1A7, Canada} \\
\r {15} {\eightit The Johns Hopkins University, Baltimore, Maryland 21218} \\
\r {16} {\eightit Institut f\"{u}r Experimentelle Kernphysik, 
Universit\"{a}t Karlsruhe, 76128 Karlsruhe, Germany} \\
\r {17} {\eightit National Laboratory for High Energy Physics (KEK), Tsukuba, 
Ibaraki 305, Japan} \\
\r {18} {\eightit Ernest Orlando Lawrence Berkeley National Laboratory, 
Berkeley, California 94720} \\
\r {19} {\eightit Massachusetts Institute of Technology, Cambridge,
Massachusetts  02139} \\   
\r {20} {\eightit University of Michigan, Ann Arbor, Michigan 48109} \\
\r {21} {\eightit Michigan State University, East Lansing, Michigan  48824} \\
\r {22} {\eightit University of New Mexico, Albuquerque, New Mexico 87131} \\
\r {23} {\eightit The Ohio State University, Columbus, Ohio  43210} \\
\r {24} {\eightit Osaka City University, Osaka 588, Japan} \\
\r {25} {\eightit Universita di Padova, Istituto Nazionale di Fisica 
          Nucleare, Sezione di Padova, I-35131 Padova, Italy} \\
\r {26} {\eightit University of Pennsylvania, Philadelphia, 
        Pennsylvania 19104} \\   
\r {27} {\eightit Istituto Nazionale di Fisica Nucleare, University and Scuola
               Normale Superiore of Pisa, I-56100 Pisa, Italy} \\
\r {28} {\eightit University of Pittsburgh, Pittsburgh, Pennsylvania 15260} \\
\r {29} {\eightit Purdue University, West Lafayette, Indiana 47907} \\
\r {30} {\eightit University of Rochester, Rochester, New York 14627} \\
\r {31} {\eightit Rockefeller University, New York, New York 10021} \\
\r {32} {\eightit Rutgers University, Piscataway, New Jersey 08855} \\
\r {33} {\eightit Academia Sinica, Taipei, Taiwan 11530, Republic of China} \\
\r {34} {\eightit Texas A\&M University, College Station, Texas 77843} \\
\r {35} {\eightit Texas Tech University, Lubbock, Texas 79409} \\
\r {36} {\eightit Istituto Nazionale di Fisica Nucleare, University of Trieste/
Udine, Italy} \\
\r {37} {\eightit University of Tsukuba, Tsukuba, Ibaraki 305, Japan} \\
\r {38} {\eightit Tufts University, Medford, Massachusetts 02155} \\
\r {39} {\eightit Waseda University, Tokyo 169, Japan} \\
\r {40} {\eightit University of Wisconsin, Madison, Wisconsin 53706} \\
\r {41} {\eightit Yale University, New Haven, Connecticut 06520} \\
\end{center}


\begin{abstract}

If the Technicolor $\omega_T$ particle exists, 
a likely decay mode is 
$\omega_T\rightarrow \gamma\pi_T$, followed by $\pi_T\rightarrow b\bar b$,
yielding the signature $\gamma b\bar b$.  
We have searched $85~{\rm pb}^{-1}$ of data collected by the CDF experiment 
at the Fermilab Tevatron 
for events with a photon and two jets, where one of the jets
must contain a secondary vertex implying the presence of a $b$ quark.
We find no excess of events above standard model expectations.
We express the result as 
an exclusion region in the 
$M_{\omega_T}-M_{\pi_T}$ mass plane.
\end{abstract}
\vspace*{0.2in}
\hspace*{1.0in} PACS numbers 13.85Rm, 13.85Qk, 14.80.-j 
\vspace*{0.2in}
\end{center}
%
%
\clearpage
\narrowtext

In the standard model of electroweak interactions, 
the elementary scalar fields of the Higgs mechanism break
electroweak symmetry and give mass to the $W$ and $Z$ gauge bosons. 
A new particle is predicted, the Higgs boson, 
which couples to quarks and leptons, and causes the fermions to acquire mass. 
Technicolor is a dynamical version of the Higgs mechanism which
does not contain elementary scalar bosons \cite{technicolor}. In
this approach, there are new heavy fermions interacting via 
the new, strong technicolor gauge interaction. These technifermions form 
vacuum condensates that perform the mass-generating functions of elementary 
scalars. They also form new boson bound states, including the $\pi_T^{0,
\pm}$, $\rho_T^{0,\pm}$ and $\omega_T$, analogous to the mesons of QCD.

In $p\bar p$ collisions a quark and an anti-quark may 
annihilate into a virtual photon which can fluctuate into a 
particle with the same quantum numbers, such as the 
hypothetical $\omega_T$.
In the model of Technicolor considered here~\cite{Lane}, the
$\omega_T$ may decay to $\gamma\pi_T$ followed by the 
decay $\pi_T\rightarrow b\bar b$; the resulting signature
is $\gamma b\bar b$.  
We have searched for events with a photon, 
a $b$ quark jet and at least one additional jet, in
$85~{\rm pb}^{-1}$ of $p\bar p$ collisions at $\sqrt{s} = 1.8$~TeV 
collected by the CDF experiment in 1994-1995.
In this Letter we describe the search
and the resulting limits on $\omega_T$ and $\pi_T$ mass combinations.

%
%

We briefly describe the relevant aspects 
of the CDF detector\cite{detector}. 
A superconducting solenoidal magnet provides a 1.4~T magnetic field
in a volume 3~m in diameter and 5~m long, containing three tracking devices.
Closest to the beamline is a 4-layer silicon microstrip detector 
(SVX)~\cite{svxnim} used to identify the secondary vertices from 
$b$--hadron decays.  Outside the SVX, a time projection chamber 
locates the $z$ position of the interaction.
In the region with radius from 30~cm to
132~cm is the central tracking chamber (CTC) 
which measures charged--particle momenta.
Surrounding the magnet coil is the electromagnetic calorimeter
which is in turn surrounded by the hadronic calorimeter.
The calorimeters are constructed of towers, subtending 
$15^\circ$ in $\phi$ and 0.1 in $\eta$, pointing to the interaction region.
In the central region (\mbox{$|\eta|<1.1$}), 
on the inner face of the calorimeter, is the 
central preradiator wire chamber (CPR).
This device is used to 
determine if a photon began its shower in the magnet coil.
At a depth of six radiation lengths into the electromagnetic calorimeter, 
wire chambers with cathode strip readout (central 
electromagnetic strip chamber, CES) measure 
two orthogonal profiles of showers.

%
%
Collisions producing a photon candidate are selected by a three-level 
trigger which requires a central electromagnetic cluster 
with $E_T>23$~GeV and limited additional energy in the region 
of the calorimeter surrounding the cluster.
Offline, we select events with an electromagnetic cluster with $E_T>$25~GeV 
and $|\eta|<1.0$.  Electron and jet backgrounds are reduced by requiring 
the cluster to be isolated from additional energy in the 
calorimeter, other energy deposits in the CES, 
and charged--particle tracks in the CTC.
These requirements yield a data sample of 511,335 events.

%
%
Photon backgrounds, dominated by jets that fragment to an energetic
$\pi^0\rightarrow \gamma\gamma$ and are misidentified as a single photon, 
are measured using the shower shape in the CES system for 
photon $E_T<35$~GeV and the probability of a conversion 
before the CPR for $E_T>35$~GeV~\cite{photons}. 
We find $55\pm 1\pm 15\%$ of these photon candidates are 
actually jets misidentified as photons.

%
%
Jets in the events are clustered with a cone of 0.4 in $\eta-\phi$ space.
The jet energies are corrected for calorimeter gaps and
non-linear response, 
energy not contained in the jet cone and underlying event energy\cite{jets}.
We then select events with at least 
two jets, each with $E_T>$30~GeV and $|\eta|<2.0$.
This reduces the data set to 10,182 events.

One of the jets is required 
to be identified as a $b$--quark jet by the algorithm used in the top--quark 
analysis\cite{top}.  
This algorithm searches for tracks in the SVX
that are associated with the jet but not associated with the 
primary vertex, indicating they come from the decay of 
a long--lived particle.  
We require that the track,
extrapolated to the interaction vertex, has a distance of closest approach
greater than 2.5 times its uncertainty.
At least two of these tracks must 
form a vertex that is displaced from the interaction vertex.
The tag's decay length, $L_{xy}$, is defined in the 
transverse plane as the dot product of the 
vector pointing from the primary vertex 
to the secondary vertex and a unit vector along the jet axis.
We require $|L_{xy}|/\sigma >3$.
These requirements constitute a ``tag''.
In the data sample the tag is required to be positive, with $L_{xy}>0$.
This final selection reduces the data set to 200 events.

%
%
A sample of multi-jet events is used to study the backgrounds 
to the tags\cite{topback}.  
For each jet in this sample, the $E_T$ of the jet,
the number of SVX tracks associated with the jet, and the scalar sum
of the $E_T$ in the event are recorded.  
The probability of tagging the jet is 
determined as a function of these variables for negative tags, 
with $L_{xy}<0$.  
Negative tags occur due to measurement resolution and errors in reconstruction.
Since these effects should 
produce negative and positive tags with equal probability,
the negative tagging probability
can be used as the probability of finding a positive tag due to 
mismeasurement (mistags).

%
%
To estimate the standard model background to the 200-event data sample,
we sum three sources: events with jets misidentified as photons,
events with photons and mistags and events with 
standard model production of photons with heavy flavor quarks.
Using the photon background method described above, 
the number of events with a jet misidentified as a photon 
is $56\pm 30\pm 8$ events~\cite{uncert}. 
The large statistical uncertainty
is due to the low discrimination power of the method, 
and the systematic uncertainty 
reflects the uncertainty in the background composition 
and the level of internal consistency in the method.
It is necessary to directly measure the jets misidentified as 
photons in this tagged sample, 
rather than apply a universal photon background fraction, since the production
of photons in events with specific quark content will have different 
production cross sections, due to the quark charges and masses, and 
therefore different ratios to the jet backgrounds.

To estimate the number of real photons and mistags,
we apply the photon background 
method and the negative tagging probability to the 
data sample before the tagging requirement.  The estimate of this background
is $27\pm 5\pm 14$ events.  
The 50\% systematic uncertainty accommodates 
a discrepancy in the number of predicted and observed 
negative tags (312 and 197, respectively) in a sample
of events with a photon and one jet.
This discrepancy may be caused by the difference in the QCD control sample,
where two jets opposite each other cause hadronization with a balanced set of 
tracks to locate the primary vertex, and the photon sample, which does 
not have the same balanced tracking topology.
The final source of backgrounds,
standard model production of a photon with a heavy quark, 
is estimated using a custom Monte Carlo~\cite{MLM}.
We expect $25\pm 2 \pm 13$ events from $\gamma b\bar{b}$  production
and $23\pm 3 \pm 12$ events from $\gamma c\bar{c}$ production.
The systematic uncertainty includes a conservative estimate of the 
uncertainty in the leading--order calculation.
The total background estimate is $131\pm 30\pm 29$ events, and 
correlations between uncertainties have been included.
We conclude that the 200 events in the data sample 
do not constitute a significant excess over standard model expectations.

\begin{figure}
\epsfxsize=11.55cm
\epsfysize=16.36cm
\vspace*{-1cm}
\hspace*{10mm}
\epsfbox{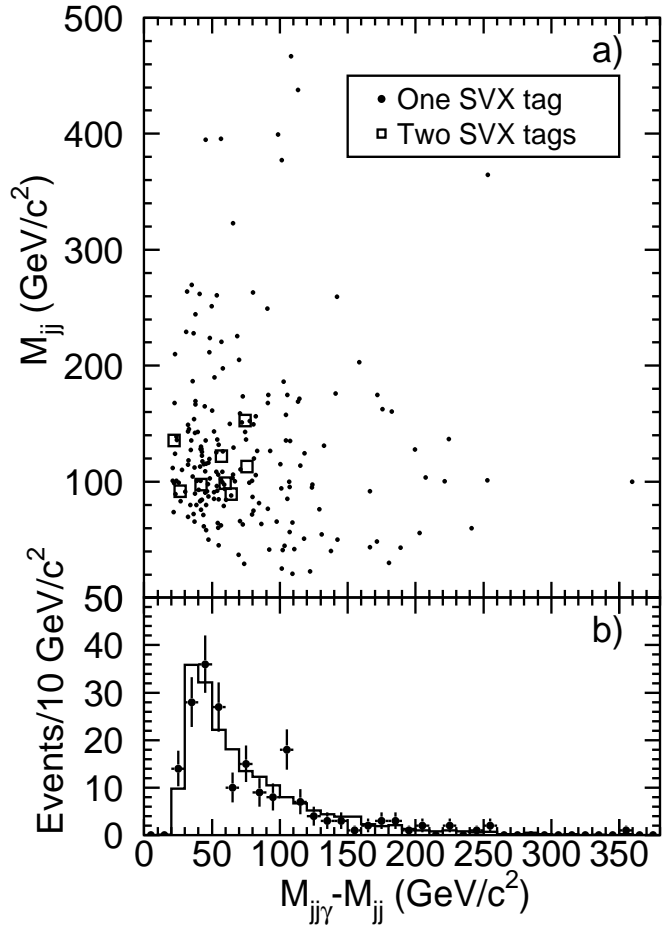}
\vspace*{-0.7cm}
\caption{a) The distribution of 
$M_{jj}$ plotted {\it vs } $M_{jj\gamma}-M_{jj}$ 
for the 200 events with a photon, a tagged jet and a second jet
in 85~pb$^{-1}$ of data. b) The projection of the same data in 
$M_{jj\gamma}-M_{jj}$.  The data are represented by the points
and the predicted background, normalized to the data, by the histogram.
A signal at the baseline model point ($M_{jj}=110$~GeV, 
$M_{jj\gamma}-M_{jj}$=100~GeV) would have a width of approximately 20 
and 5~GeV in these variables respectively. }
\label{fig:scat}
\end{figure}

%
%
To set limits on the Technicolor model\cite{Lane}, we investigate 
points in the $M_{\omega_T}-M_{\pi_T}$ mass plane on a 20~GeV/c$^2$ grid.  
We also investigate a baseline point at 
$M_{\omega_T}=$210~GeV/c$^2$ and $M_{\pi_T}=$110~GeV/c$^2$,
 suggested by the authors of the model.
At each point in the grid, limits are set using the following procedure.
Two invariant masses are calculated for each event: the mass of the 
tagged jet and the highest-$E_T$ untagged jet,
$M_{jj}$, corresponding to the $\pi_T$ mass, and the mass of the two jets
plus the photon, $M_{jj\gamma}$, corresponding to the $\omega_T$ mass.
The distributions of $M_{jj}$ and $M_{jj\gamma}-M_{jj}$ for the data 
are shown in Figure~\ref{fig:scat}.
We collect the events with $M_{jj}$ within a window around $M_{\pi_T}$, 
$|M_{jj}-M_{\pi_T}|<0.36M_{\pi_T}$, which is selected to 
be 90\% efficient for the signal.  For these events, 
we histogram the mass difference $\Delta M=M_{jj\gamma}-M_{jj}$
where a signal would appear as a peak.  
The mass difference has good resolution since 
the poor jet resolution is largely canceled.
We fit the $\Delta M$ spectrum above 50~GeV/c$^2$
to a background distribution and 
a Gaussian peak.  The central value of the Gaussian is fixed to 
the $\Delta M$ of the grid point 
and the width is fixed to the expected value, 
\mbox{$\sqrt{\Delta M/(4~{\rm GeV/c^2})}$~GeV/c$^2$},
which is the experimental resolution.
(The Technicolor particles have negligible natural width.)
The fit likelihood is convoluted with the systematic 
uncertainty (described below) and integrated to find the 95\% confidence
level limit on the number of signal events.
The fit is performed twice, with different background functions
(an exponential and a sum of two exponentials).
Little difference in the fit results is observed, and 
the result which leads to the more conservative limit is used. 
All fits report less than a $2.2\sigma$ excess.

%
%
Efficiencies are measured using the 
{\tt PYTHIA 6.1} \cite{PYTHIA} Monte Carlo program and
a detector simulation.  For the masses at the baseline point, 
$M_{\omega_T}=$210~GeV/c$^2$ and $M_{\pi_T}=$110~GeV/c$^2$, 
we find 44\% of generated events contain a photon with $E_T>25$~GeV
and $|\eta|<$1.0.
The efficiency of the photon trigger, fiducial cuts and 
identification cuts, 
calibrated with electromagnetic clusters in $Z\rightarrow e^+e^-$ events, 
is 59\%.  
The efficiency for at least one jet with $E_T>30$~GeV
and $|\eta|<$2.0 is 91\%.
The efficiency for one or more 
tag, calibrated with $b\bar{b}$ data, is 36\%.
After including the efficiency for the second jet (66\%) and the 
$M_{jj}$ mass window cut (90\%), the total \mbox{A$\cdot\epsilon$} 
(where A is the acceptance and 
$\epsilon$ is the efficiency) for this choice of masses is 5.1\%.
The maximum \mbox{A$\cdot\epsilon$} of approximately
10\% is obtained at the largest $M_{\pi_T}$ and $\Delta M$.

%
%
The combined systematic uncertainty of 22\% consists of contributions
from the uncertainties in photon identification efficiency (14\%),
tagging efficiency (9\%),
luminosity measurement (8\%),
initial--state and final--state radiation (6\%),
jet energy scale (6\%),
parton distribution function (5\%),
and Monte Carlo statistics (4\%).

Finally, \mbox{A$\cdot\epsilon$}, 
the luminosity (\mbox{85~pb$^{-1}$}), and 
the upper limit on the number of observed events 
are combined to yield a limit
on \mbox{$\sigma\cdot$B} (cross section times branching ratio)
which can be compared to the theoretical prediction.
The model parameters we use are $N_{TC}=4$ 
(the number of technicolors, analogous to the three colors in QCD), 
$Q_D=Q_U-1=1/3$ (the techniquark charges),
and $M_T=100$~GeV/c$^2$ 
(a dimensionful parameter of order the technicolor interaction scale; 
cross section scales roughly as $M_T^{-2}$).

\begin{figure}
\epsfxsize=8.25cm
\epsfysize=10.8cm
\vspace*{-1cm}
\hspace*{10mm}
\epsfbox{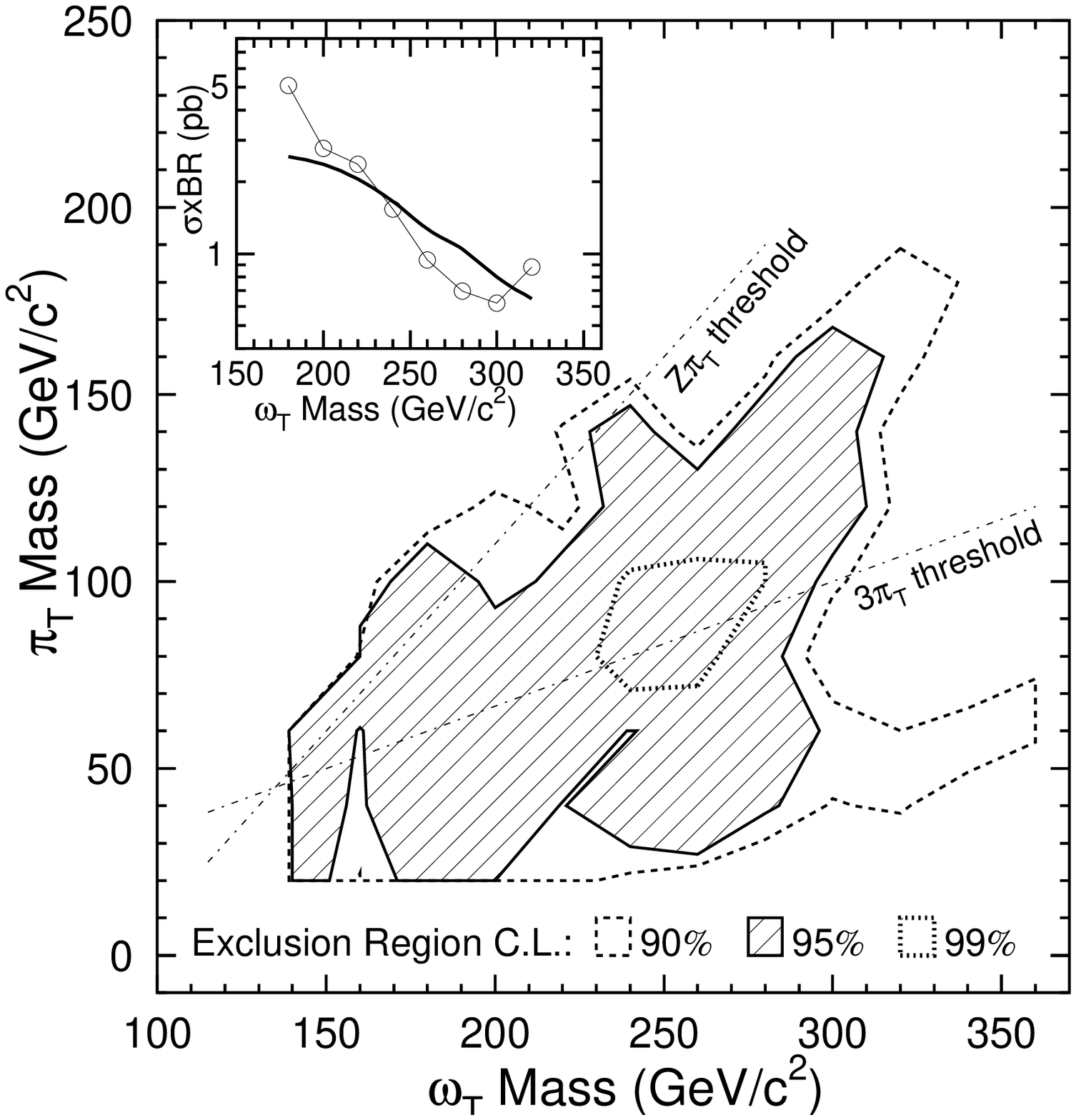}
\vspace*{-0.7cm}
\caption{
The 90\%, 95\% and 99\% confidence level 
exclusion regions in the $M_{\omega_T}-M_{\pi_T}$ mass plane
for $\omega_T\rightarrow \gamma\pi_T$ followed by $\pi_T\rightarrow b\bar{b}$.
The integrated luminosity of the data sample is 85~pb$^{-1}$.
The difference between the regions is an indication of the robustness
of the limits. In the regions below the dash--dotted lines, 
additional decay modes become available but are assumed to be negligible.
The inset shows the limit on \mbox{$\sigma\cdot$B} for 
$M_{\pi_T}=120$~GeV/c$^2$.  The circles represent the limit and the
solid line represents the theoretical prediction.
}
\label{fig:scatlim}
\end{figure}

With this parameter set, the $\omega_T\rightarrow \gamma\pi_T$
branching ratio ranges from 35\% to 85\%
in the regions we investigated (the $\pi_T\rightarrow b\bar{b}$
branching ratio is assumed to be 100\%).
The competing $\omega_T$ 
branching ratios are $q\bar q$, $\ell^+\ell^-$ and $\nu\bar \nu$.
The $Z\pi_T$ and
$3\pi_T$ branching ratios have not been calculated and are 
assumed to be negligible in making the theoretical predictions.   
We use the CTEQ4L
parton distribution function in the Monte Carlo generation.  
The leading--order theoretical cross section has been 
scaled up by a K--factor of 1.3, which corrects for higher-order 
diagrams and is derived by comparing the 
{\tt PYTHIA} Monte Carlo $Z$ cross section to the cross section 
measured in the CDF data.
With these model assumptions, combinations of 
$\omega_T$ and $\pi_T$ mass can be excluded at the 95\% 
confidence level as shown in Figure \ref{fig:scatlim}.
To define the 
exclusion region we interpolate between the grid points.  The fits to 
$\Delta M$ are sensitive to fluctuations in the data and this 
causes the ragged exclusion region boundary.
For the baseline point, the model predicts a branching ratio 
of 63\%, and a $\sigma\cdot$B of 2.32~pb (10.2 events expected).
At this point, a $\sigma\cdot$B of more than 2.38~pb (10.4 events) is excluded
at the 95\% confidence level.

%
%

In conclusion, we have searched for the production of a Technicolor $\omega_T$
that decays $\omega_T\rightarrow \gamma\pi_T$
followed by the decay $\pi_T\rightarrow b\bar{b}$ in \mbox{85~pb$^{-1}$}
of data collected in the CDF experiment.  
We observe no evidence of this production and 
exclude a significant range
of $\omega_T$ and $\pi_T$ mass combinations.

%
%

     We thank the Fermilab staff and the technical staffs of the
participating institutions for their vital contributions.  
We would like to thank K. Lane 
for useful discussions and J. Womersley for 
help with the Technicolor Monte Carlo program.
This work was
supported by the U.S. Department of Energy and National Science Foundation;
the Italian Istituto Nazionale di Fisica Nucleare; the Ministry of Education,
Science and Culture of Japan; the Natural Sciences and Engineering Research
Council of Canada; the National Science Council of the Republic of China; 
the Swiss National Science Foundation; and the A. P. Sloan Foundation.

\clearpage

\begin{thebibliography}{99}
\bibitem{detector} F.~Abe {\it et al.},
Nucl. Instrum. Methods Phys Res., Sect. A {\bf 271}, 387 (1988).
The  $z$ axis points along the proton beam axis;
$r$ is the distance from the beamline and $\theta$ is the angle
from the $z$ axis.  Pseudorapidity  ($\eta$) is 
$\eta\equiv-$ln(tan($\theta$/2)) and 
transverse energy is defined as $E_T ={\rm E}\sin(\theta)$.
\bibitem{uncert} 
In this Letter when two uncertainties are reported, the first is 
statistical and the second is systematic.
\bibitem{MLM} The $\gamma b\bar{b}$ and $\gamma c\bar{c}$
Monte Carlo programs were provided by M. Mangano.  The calculation
is leading order based on $q\bar q$ and $gg$ initial states
and a finite $b$--quark mass.  The $Q^2$ scale is taken to be the 
photon $E_T$ squared plus the $b\bar{b}$ or $c\bar{c}$ mass squared.
A systematic uncertainty of 30\% is found by scaling $Q$ by a factor of two
and the quark masses by 10\%.  An additional 20\% systematic uncertainty 
allows for unmeasured effects.
\end{thebibliography}
\end{document}